# A Polymer Blend Substrate for Skeletal Muscle Cells Alignment and Photostimulation

V. Vurro[1,2], A. D. Scaccabarozzi[1], F. Lodola[1], F. Storti[1,2], F. Marangi[1,2], A. M. Ross[2], G.M. Paternò[1], F. Scotognella[1,2], L. Criante[1], M. Caironi[1] and G. Lanzani[1,2]*

[1]Center for Nano Science and Technology @PoliMi, Istituto Italiano di Tecnologia, Via Pascoli 70/3, 20133, Milan, Italy.

[2]Department of Physics, Politecnico di Milano, Piazza Leonardo da Vinci, 32, 20133, Milano, Italy

**Abstract**

Substrate engineering for steering cell growth is a wide and well-established area of research in the field of modern biotechnology. Here we introduce a micromachining technique to pattern an inert, transparent polymer matrix blended with a photoactive polymer. We demonstrate that the obtained scaffold combines the capability to align with that to photostimulate living cells. This technology can open up new and promising applications, especially where cell alignment is required to trigger specific biological functions, e.g. generate powerful and efficient muscle contractions following an external stimulus.

**Introduction**

In tissue engineering, cell-substrate coupling has a crucial role. Its mechanical, chemical, physical and morphological features are decoded by cells as stimuli that strongly affect their behavior. Often tools such as electrodes are further added to the cell culture in order to probe or actively stimulate the cellular processes. The use of electrodes has large practical success but it does bear a number of downsides. Alternatively, photostimulation provides advantages such as better spatial and temporal resolution, low toxicity and invasiveness. In addition, the stimulation map throughout the sample can be changed easily without the need of rewiring nor the limit of discrete points, owing to the wireless architecture of this optical strategy. The use of light to control living cells and organisms has a long history, but it is regaining interest due to the emerging of new techniques at the frontier between photonics and biology. Optogenetics,[1] NIR[2] and phototransducers based[3] stimulation are the most popular and promising techniques. In particular, the use of non-genetic phototransducers has the highest translational potential. The phototransducers may come in a variety of shapes and structure, such as plasmonic nanodots,[4,5] inorganic semiconductors devices,[6–8] nanostructures,[9–12] organic semiconductors[13–17] and small molecules[18,19]. Thanks to the similarity in composition and structure with biomolecules, organic semiconductors and molecular compounds express high biocompatibility and multifunctional actuation. Their peculiar photophysical properties allow the employment of different photostimulation mechanisms (i.e. electrical, mechanical or thermal) and at the same time, their mechanical flexibility favors conformational matching with biological environment. These properties make organic materials suitable for *in vivo* applications, where a soft yet efficient interaction with the biological counterpart is extremely important.[20,21] Furthermore, by using wet processes at room temperature, organic semiconductors are easily fabricated into a great number of forms whose morphology can mimic those of the extracellular matrix and improve both cell growth and cell coupling at the abiotic/biotic interface.[22,23] In particular, in muscular tissue engineering, in order to efficiently reproduce the natural muscle organization and achieve contraction ability, cell alignment on a suitably stiff substrate is required. This can be done by depositing extra matrix proteins[24–27] or polymer fibers that produce an aligned path to force cell's orientation.[28–30] Alternatively, a simpler approach relies on the fabrication of microgrooves onto a substrate that physically confine the cell.[31–36] There are no universal rules that define which pattern can lead to an optimal morphological stimulation, since it strongly depends on the target cell. Nevertheless, some guidelines can be identified, regarding both the width and depth of the pattern. Indeed, the depth should be ideally higher than the cell's height or at least of the order of micron.[37] The width can spun over different values depending on the application and the target cell. As a general rule in order to align a single cell, the width of a pattern should be lower than the cell's length that can be several times its height.

In this work we present a polymer substrate engineering designed to perform a double functionality: it drives cell alignment and allows the photostimulation of cell behavior. We chose Poly-(3-hexylthiophene-2,5-diyl) (P3HT), a highly studied polymer, as photo-active material. Upon blending it with high density polyethylene (HDPE)[38] we obtained a free standing polymer film with good mechanical properties that preserves P3HT optoelectronic features. By using a maskless laser ablation approach we were able to inscribe a customizable guiding pattern in the films suitable for achieving aligned cell growth. The combination of blending and micromachining is used as a novel approach to achieve multifunctional substrate without the employment of complex and expensive fabrication methods such as photolithography. Moreover, the upscaling of this approach can be easily achieved in order to increase the complexity of the realized pattern. As a simple model system for preliminary biocompatibility and photostimulation studies we used Human embryonic kidney (HEK) cells. Once parameters had been optimized we focused on to C2C12, a murine myoblast cell line characterized by specific biophysical parameter, like myotube diameter ranging from 10 μm to 25 μm and length ranging from 130 μm to 520 μm. [39] C2C12 show many peculiar properties of skeletal muscle cells (i.e. the ability to differentiate toward multinucleated myotubes that can display contraction) thus making them a widely accepted *in vitro* model to validate our purpose.

**Result and Discussion**

It has been widely reported in literature that P3HT can be blended with a large variety of insulating polymers preserving its optoelectronics properties.[40–44] The blend strategy is employed to introduce desired processing and mechanical properties of commodity polymers (i.e. common plastics) in organic electronics applications. Plastic "additives" are ductile, display high elongation at break, while in contrast semiconducting polymers are typically rather brittle owing to their low molecular weight.[43] Moreover, insulating polymers have been reported to be beneficial in a number of applications leading for instance to improved charge carriers mobilities, increased device stability,[45–47] tailored optical properties,[48,49] decreased energetic disorder. [50] We focus here on HDPE, a semi-crystalline insulating polymer that has been already employed in combination with P3HT to produce both thin and thick films for a large variety of applications. Remarkably, the blend system retains semiconducting optoelectronic properties even at very low concentrations of the photoactive material in the insulating matrix, while showing improved mechanical properties. Indeed, the improved viscoelastic properties of the blend system allow the production of robust self-standing films that can be used directly in tissue engineering, while films comprising neat semiconducting polymers would be too brittle to do so. Beside the combination of mechanical and optoelectronic properties, blending has an economic advantage for HDPE being much cheaper than P3HT, which is particularly useful when producing thick films (tens of microns).

**Fabrication Process**

The fabrication process is designed to obtain a photo-excitable substrate suitable for laser pattern ablation. We assumed that a surface pattern would be able to induce a satisfactory cell alignment if it had a characteristic depth higher than the cell diameter and a width shorter than the cell length, as we discussed in the previous section (for a muscle cell height ≈ 15 μm and length ≈ 200 μm). [39] Following these requirements, we selected a pattern geometry formed by parallel lines with a width of 60 μm and a depth of at least 15 μm, thus requiring a film thickness >20 μm. The fabrication process (**Figure 1**) is based on a pre-compounding step performed in solution, followed by solid-state processing of the self-standing film and final laser patterning. In principle, the solution processing could be avoided and the polymer:polymer compounding could be achieved upon melt mixing P3HT and HDPE, leading to a solvent free process. However, the latter requires the employment of a large volume of materials and high temperature of processing owing to the high melting point of P3HT (ca. T = 238°C).[51] Consequently, the solution-based method is more practical. Once a homogeneous solution is achieved, it is drop cast to obtain a solid state deposit. Despite this deposit exhibits the required self-standing and mechanical properties, it lacks the necessary thickness and homogeneity. In order to reach a suitable film thickness, around 30 μm, we delaminated, folded and pressed the polymer blend into a uniform film. The blend strategy is particularly useful, not only to tune the mechanical properties of the films, but also to control the optical density. Indeed, HDPE does not absorb in the UV-vis range (**see Figure S1**) and when intermixed with the semiconductor it enhances the transparency of the otherwise opaque P3HT thick film. Films obtained with a concentration of P3HT in the range 1-5 wt% show an optical density suitable for photostimulation experiments at the required thickness of 30 μm.

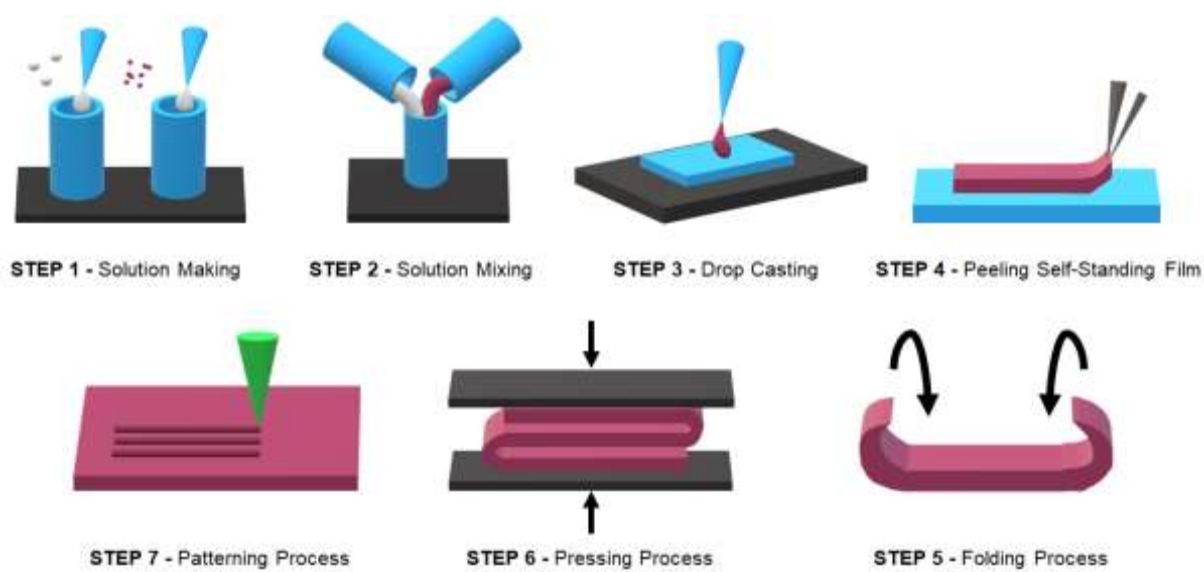

*Figure 1. Schematic representation of the multistep fabrication process.* *Step 1 and 2 regards the solution making process: two different solutions are made for each neat polymer dissolved in the same solvent and at the same concentration and successively mixed with a specific ratio. From step 3 to step 6 the realization of a thick HDPE:P3HT film is illustrated. The process is based on drop casting (step 3), peeling and subsequent folding of the solid deposit (step 4-5) that precede the pressing step (step 6). Finally, a patterning process is performed upon laser ablation (step 7). In the sketch are show a microscope slide glass in blue used in the deposition step and a metal hot plate in dark gray used in deposition and pressing step.*

**Photophysical Properties**

The optical properties and elementary excitation dynamics of the P3HT:HDPE films were investigated by measuring absorbance, photoluminescence (**Figure S1**) and transient photoinduced absorption (**Figure 2**) in a spin-cast P3HT thin film (P3HT-TF) and for three different HDPE:P3HT pressed blends (with P3HT concentration equal to 1, 3 and 5 wt%). Note that HDPE is virtually transparent in the considered spectral range (400 - 800 nm) as shown in **Figure S1**, thus spectroscopy essentially addresses P3HT properties. [52–56] The optical properties of the three blend samples are very similar to that of the neat film suggesting that: a) the fabrication process has a minor effect onto the P3HT photophysics and b) P3HT successfully phase separates from HDPE in the blend (suggested by the highly visible vibronic structure). Indeed, it has been well established how the processing conditions, and in particular the deposition temperature, play a critical role in insulating:semiconducting polymer blends not to hamper the solid state optoelectronic properties of the semiconductor.[57–59] Here, we selected a deposition temperature of 100 °C, in a range in which P3HT is known to phase separate into micro-crystallites prior the HDPE crystallization during the solvent evaporation, leading to an unperturbed UV-vis absorption of the semiconductor. The ordered nature of P3HT is confirmed by the highly structured absorption spectra, typical of crystalline aggregates of this polymer. By and large the transient absorption data confirm this picture, even though quantitative differences are present depending on the blend composition. In particular, in the 1% blend the photoinduced absorption (PA) band ($\Delta T < 0$) appears much wider than in the other spectra, suggesting a more efficient charge separation process. In the 3% and 5% samples the first vibronic resonance of the bleaching band ($\Delta T > 0$) is enhanced suggesting a higher degree of order in the P3HT domains. This hypothesis is enforced by the red shift of the emission spectra as a function of the P3HT concentration that suggests an enhanced aggregation as well as a planarization of the P3HT molecules (**Figure S1**). Despite a more detailed analysis is beyond the scope of this work, we can speculate that the crystal packing of P3HT chains secluded in the blend might favor singlet fission into triplets. Overall, we conclude that P3HT properties in the blend are similar to those in the neat film, possibly with higher degree of planarization that enhances inter chain coupling and in turn the crystal versus amorphous contribution.

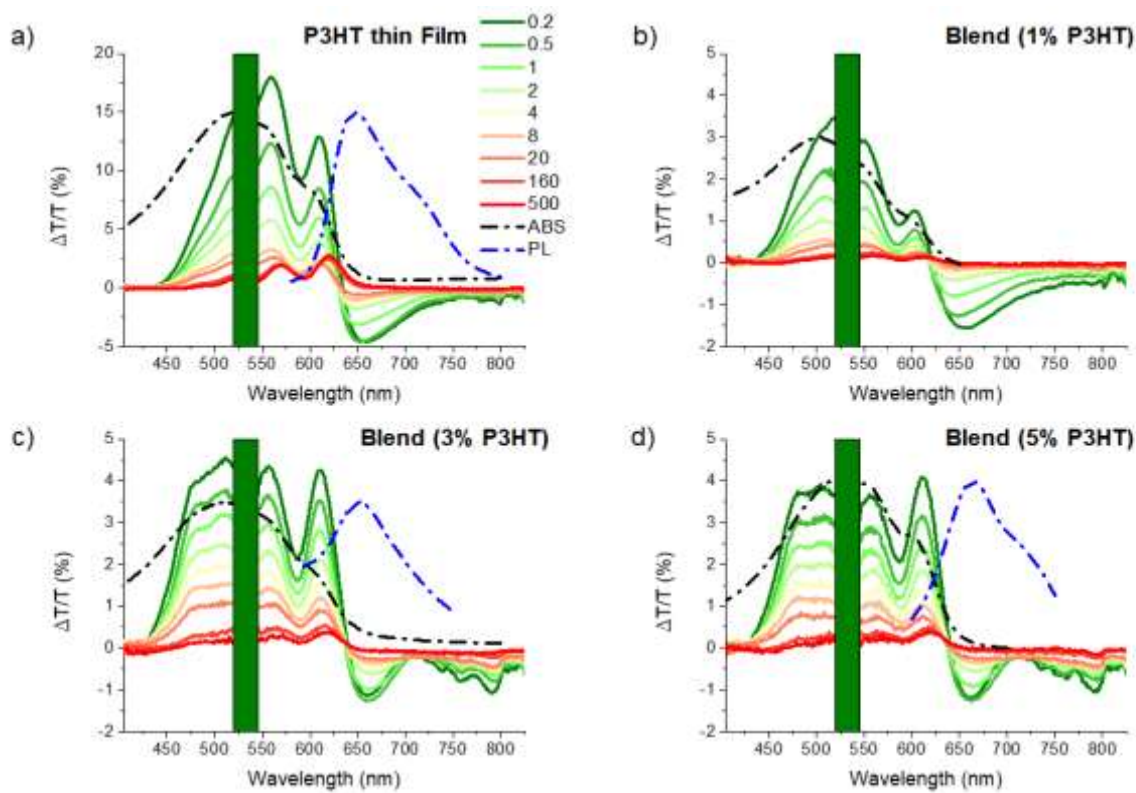

*Figure 2. Comparison between spectroscopic behavior of P3HT-TF and the blended film obtained with different percentage of P3HT (a- P3HT-TF, b- blend (1% P3HT), c- blend (3% P3HT), d- blend (5% P3HT)). In each panel the absorbance spectrum (dash-dot black line), the photoluminescence spectrum (dash-dot blue line) and the transient absorption spectra (solid lines) are reported. The time evolution is indicated by the color gradient of the line (from dark green = 0.2 ps to dark red = 500 ps). In 2-b the photoluminescence spectrum is not reported because it is buried underneath the dominant scattering feature of the sample (see Supplementary Information for more details).*

**Substrate Biocompatibility**

The biocompatibility of the substrate and of the patterning process were tested with two distinct assays. For substrate biocompatibility the alamarBlue proliferation test was performed with HEK cells (**Figure 3b**). These cells were preferred for this kind of methodology due their dimension, which is lower than C2C12 and therefore more suitable for acquiring data for longer proliferation times and for choosing a higher seeding density, a crucial aspect for this test. The biocompatibility of the patterning process was also tested, to be sure that laser ablation did not produce toxic fabrication debris. Conversely, C2C12 viability was evaluated with the HOECHST 33342/NucGreen Dead 488 ReadyProbes assay (**Figure 3c**). Both the film and the patterning process highlight very good biocompatibility showing proliferation and viability rates similar to those obtained seeding the cells on glass, that represents our standard control condition.

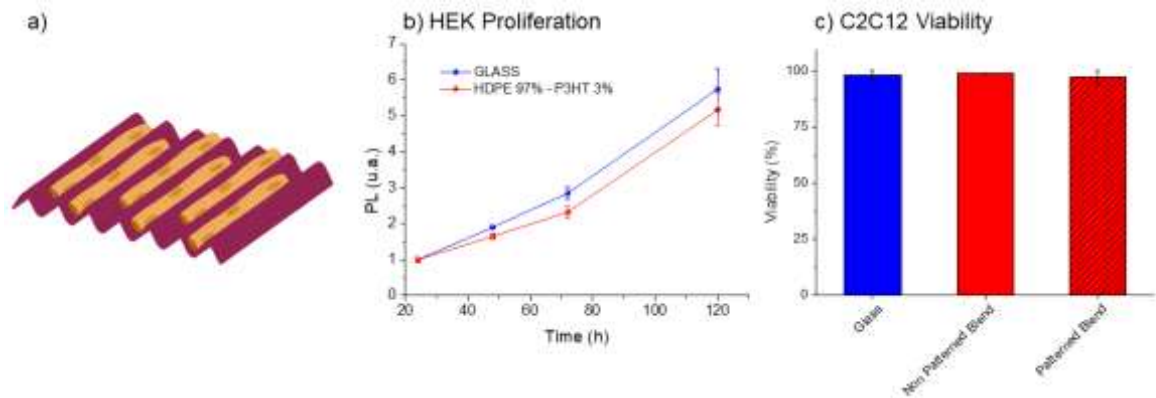

*Figure 3. Results of Biocompatibility Tests. a- sketch of cells seeded on the patterned blend film. b- Proliferation assay (alamarBlue) performed on HEK cells. Measurements were performed after 1, 2, 3 and 5 days after seeding. c- Viability test for C2C12 cells cultured on bare glass, blend and patterned film (3% P3HT). Data are presented as mean ± s.e.m., n = 3 wells measured 5 times for proliferation and n = 80 cells for viability.*

**Cell Photostimulation**

Having established that the blend films retain their optoelectronic properties and have suitable biocompatibility, we went on studying the performance in cell photo-stimulation. The electrophysiological properties were measured by carrying out patch clamp experiments in whole-cell configuration and current clamp mode ($I = 0$). Using this approach, changes of the cell membrane potential are recorded during light excitation. As we showed in previous works, [60,61] the cell membrane undergoes a depolarization and a subsequent hyperpolarization upon illumination when cells are plated on P3HT-TF. We measured the cell membrane variation of both HEK and C2C12 cells seeded on thick solid state pressed P3HT:HDPE films when short (20ms) and long (200ms) light pulses ($\lambda$=530 nm) are applied (**Figure 4**) at a fixed power density (40mW/mm$^2$).

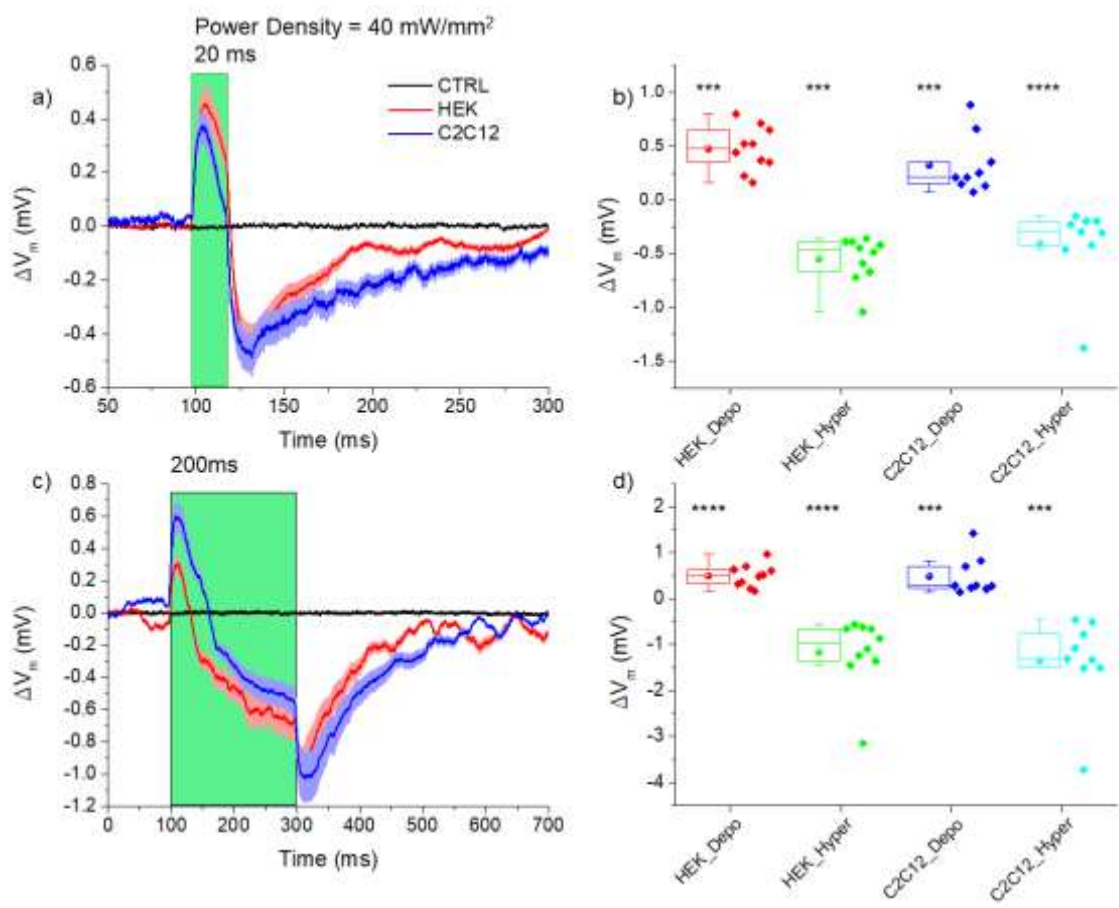

*Figure 4. Electrophysiological Response of HEK and C2C12 cells.* *Representative curves of the membrane potential during light excitation for 20 ms (a) and for 200 ms (c) light pulses. The black curves are the control measurement from cells grown on HDPE films. Box plots of depolarization (b) and hyperpolarization (d) changes obtained for both HEK (n=10) and C2C12 (n=10) cells with a photoexcitation density of 40 mW/mm$^2$ at 530 nm. Statistical comparison is evaluated between resting and light induced membrane potential value.*

We detected a transient depolarization upon illumination with short pulses, quickly evolving into hyperpolarization when the light was switched off. The membrane potential recovered its resting value approximately after 300 ms (**Figure 4a**). By using longer pulses the initial depolarization turned into hyperpolarization during illumination. At the light offset there was again an overshoot increase of the hyperpolarization, followed by a recovery to the initial resting potential in about 300 ms (**Figure 4b**). A very good reproducibility of the data is observed, as it can be noticed from the low dispersion of the data in both the cell lines and the pulse length shown in **Figure 4 b-d**. We performed measurements on a control sample (black line **Figure 4**) without observing any signal during measurement, ruling out a direct interaction between cell or HDPE with light. Once the photoinduced effect has been demonstrated, we performed further studies on the biophysical mechanisms. In previous works, the photoinduced membrane potential variation, measured in cells seeded on bare P3HT films and subjected to light stimulation protocol similar to the one adopted here, was assigned to a thermal effect. In particular, the initial depolarization is reported to be due to the increase of capacitance of the cell membrane, while the following hyperpolarization, still under light exposure, is assigned to the change in the cell baseline.[60] The dark evolution towards equilibrium is ascribed to the recovery of capacitance and cooling off. In principle our blend could have a different thermodynamic behavior under illumination when compared with the neat thin film.

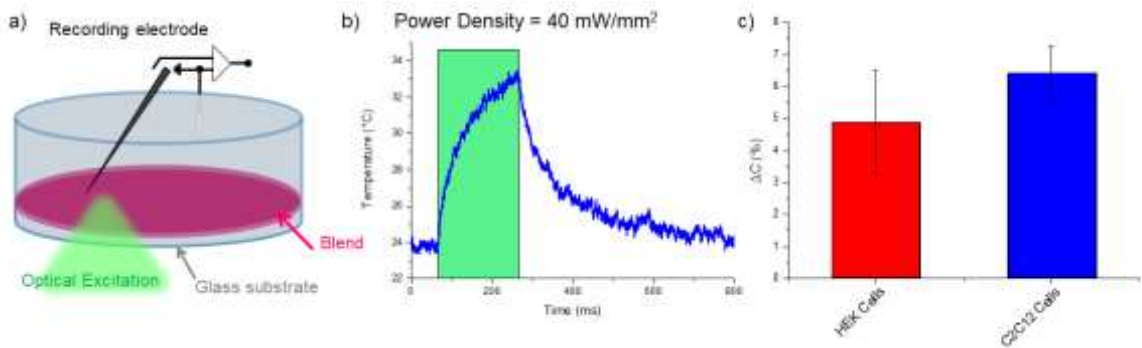

*Figure 5. Proposed photothermal mechanism. a- Sketch representation of the calibrate pipette method. b- Temperature variation of the extracellular bath upon 200 ms photoexcitation as a function of time. The experiment is performed at room temperature (24 °C) with a photoexcitation density of 40 mW/mm² at 530 nm. c- Light induced variation (mean ± s.e.m.) of the whole cell capacitance for HEK (n = 10) and C2C12 (n = 10) cells.*

We therefore measured the temperature variation in close proximity of the substrate, using the calibrated pipette method (**Figure 5, a-b**) described by Martino *et al.*[54] Effectively, upon irradiation with a 200 ms light pulse, the temperature rose up to 33°C, with $\Delta T_{Blend} = +9°C$, which is comparable to that observed on neat P3HT thin films, $\Delta T_{P3HT-TF} = +7°C$ with a comparable power density (≈40

mW/mm$^2$). Apparently, absorption in the blend is large enough to give rise to a change in temperature that slightly overcomes that observed in the neat film, despite the low P3HT density in the blend. This could probably be due to the different thickness and the different thermal conductivity of the two films. To confirm the mechanism reported above we also measured the capacitance in dark condition and upon light excitation by applying the double sinusoidal technique. **Figure 5-c** reports the results. As expected, the capacitance increases due to light excitation in both the investigated cell lines. The percentage increase, ΔC, is 4.87 ± 1.62% for the HEK cells and 6.40 ± 0.85% for C2C12 cells.

**Laser Ablation and Cell Orientation**

Once the cell photostimulation ability of the plain pressed blend was assessed, we further aimed at micromachining its surface in order to force cell alignment. The manufacturing process creates a grooved line pattern on our blend substrate taking advantage from the ultra-short pulse laser direct surface ablation technique. This innovative and maskless fabrication technique gives a high degree of freedom in designing both the shape and the wall dimension of such patterns, allowing geometries that are more complex. By changing the repetition rate, the pulse energy or the position of the laser beam focal spot it is possible to tune the wall depth, the geometry (which can also be curvilinear) and control the spatial resolution of the chosen pattern (as shown in **Figure S3**). Furthermore, direct laser ablation provides a quicker and cheaper manufacturing technique, compared to classical lithographic methods, that well matches the cost-effectiveness of the blended substrates presented in this work. We selected a simple square cross-section target pattern with a wall depth greater than 20 μm and a width of 60 μm (**Figure 6**). We then assessed the ability of the realized pattern to align cells with Scanning electron microscopy (SEM), by acquiring images of C2C12 cells seeded directly on the substrates. To better visualized the effect of the pattern we first selected the experimental conditions that allowed the growth of isolated cells. A representative image acquired in this condition is shown in **Figure 7-a**, where it is possible to appreciate how the substrate pattern effectively steers the cells orientation. In **Figure 7-b** we show the polar plot of the alignment angles, where 0° corresponds to the pattern direction.

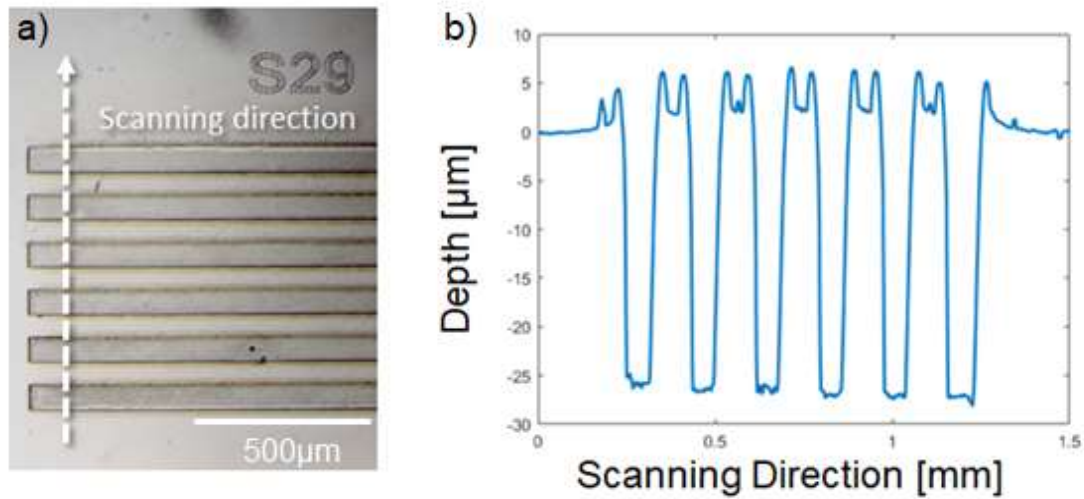

*Figure 6. Obtained Pattern by Laser Ablation.* *a- Microscope image of the pattern obtained with the utra-short pulse laser direct ablation process. b- A profile of the obtained pattern shows its characteristic dimensions, a depth of 25 μm and a width of 65 μm. The scanning direction is represented with a dashed arrow in panel a.*

The mean orientation angle is 12.95°±1.52°. Moreover, a good alignment efficiency is observed, indeed 77% of the cell's directions are included in the angular interval 0°-15° and 90.3% in 0°-30°. Having established that cell growth anisotropy is achieved in the low-density cell regime, we carried out also high cell density experiments trying to achieve a muscular tissue (see supporting materials).Despite it is hard to obtain a quantitative evaluation of the alignment; nevertheless it is still possible to observe that the cell growth follows the pattern direction (see **Figure S5**). The effect is particularly evident at the edge of the pattern where is possible to appreciate an evolution of the cells from a random to an aligned orientation condition canalized by the pattern.

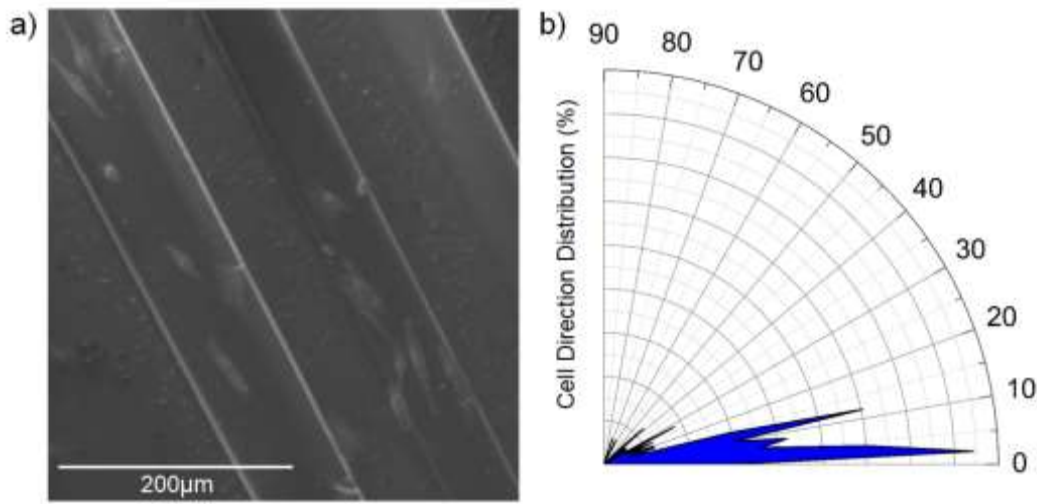

*Figure 7. Pattern Induced Cell Orientation. a- SEM image of C2C12 seeded on patterned substrate in isolated cells condition. b- polar plot of the cell directionality distribution normalized on the pattern direction (n = 83 cells)*

**Conclusion and Perspective**

Substrate mediated cell engineering for both growth control and photostimulation is a wide and well-established area within biotechnology research. Even though the two sub-fields share common application paths and are focused on a similar biophysical interaction (cell-substrate coupling), they are normally studied separately by researchers with different expertise. Here we combine, for the first time, the ability to align and photostimulate cells within the same substrate. Thanks to the features of the fs-micromachining facility it is possible to obtain patterning of any shape and size, without mask, easily and quickly adaptable to each series of different cells. The same setup also allows a high-resolution soft-cut of the material to give a macro customized shape to the patterned areas, if and when necessary. This is an enabling technology that can open up new applications, especially when cell alignment is mandatory to carry out specific functions, e.g. muscle contraction. We demonstrated both successful alignment and photostimulation. Ideally, exploiting this novel and multitasking approach will permit to fabricate systems in which fully developed skeletal or cardiac muscle cells can grow in an organized array that allows carrying out specific functions under light control. Further developments of this work regard replacement of both the photoactive and the insulator material. The former aims to move the optical absorption spectra towards wavelengths that can penetrate deeper into the tissue in order to broaden the spectrum of possible biological applications. The insulator instead can be replaced with hydrogel or elastomeric material to well reproduce biological stiffness. Moreover, the obtained blend, in its solution step, and the mentioned possible ameliorations could be optimized to exhibit ink-properties compatible with 3D-printing techniques, thus aiming to reproduce the structural complexity of living tissue.

**Experimental Section**

**Sample Preparation**

Regio-regular P3HT (99.995% purity, 20000–45000 molecular weight) was purchased from Sigma Aldrich and used without any further purification. High Density Polyethylene (125000 molecular weight) was purchased from Alfa Aesar and used without any further purification.

Solutions of the two material (P3HT and HDPE) were prepared in di-chlorobenzene at concentration of 15 g/l, mixed in a specific ratio (1-99%; 3-97% and 5-95%) and stirred for at least 30 minutes. The obtained mixing solution was drop casted on a hot plate at 100°C. Once the solvent was evaporated the obtained film was folded and pressed with a hydraulic press with several step to arrive to the maximum pressure of 7 MPa held for 8 minutes. Although the solution step is not required and the exact amount of powder can be directly pressed to obtain the film, in this work we decided to go through solution to be more precise in the mixing step. The obtained film was circular with a diameter of 18 mm and thickness of 30 μm.

**UV-Vis Absorption**

The UV-Vis absorption spectra were acquired with a Perkin Elmer Lambda 1050 spectrophotometer equipped with deuterium (280-320 nm) and tungsten (320-3300 nm) lamps. The signal is acquired with three detectors acting in different spectral region (photomultiplier [180, 860] nm, InGaAs [860, 1300] nm and PbS [1300, 3300] nm.

**PL measurements**

The PL measurements were carried out with a Horiba Nanolog Fluorimeter equipped with two detectors (photomultiplier and InGaAs). The sample was excited at 530 nm using the same LED lamp and power employed for transmission measurements (see above).

**Ultrafast time-resolved spectroscopy**

For the femtosecond TA measurements, we used a Ti:Sapphire laser with a repetition rate of 1 kHz and a pulse width of 100-150 fs. We excited the films at 530 nm. The beam was generated by means of an optical parameter amplifier and probed with a white light beam generated by a $CaF_2$ plate. The excitation energy was of ≈ 30 µJ and the beam spot size of ≈ 200 µm in diameter.

**Laser Pattern**

The micromachining setup consists of an amplified Yb:KGW femtosecond laser system (Pharos, Light Conversion) with 230 fs pulse duration, 515 nm wavelength (frequency doubled), 500 kHz repetition rate focused with a 20X - 0.42 NA microscope objective (Mitutoyo). Computer-controlled, 3-axis motion stages (ABL-1000, Aerotech) interfaced by CAD-based software (ScaBase, Altechna) with an integrated acousto-optic modulator are used to translate the sample relative to the laser irradiation desiderate patch. An average power (pulse energy) of 15 mW (30 nJ) and a scan speed of 1 mm/s are used to laser pattern the substrates.

**Cell culture maintenance**

HEK and C2C12 cells were cultured in T-75 flasks containing Dulbecco's modified Eagle's medium (DMEM) supplemented with Fetal Bovine Serum (10%), glutamine (2 mM $L^{-1}$), Penicillin (100 IU $ml^{-1}$) and Streptomycin (100 µg $ml^{-1}$). Culture flasks were maintained in a humidified incubator at 37°C with 5% $CO_2$ and, when at confluence, plated at a density of 15.000 cells $cm^2$ and cultured for 48 h on samples. Prior to cell plating a layer of fibronectin (2 µg $ml^{-1}$ in PBS buffer solution) was deposited on the samples surface and incubated for 1 hour at 37°C, to promote cellular adhesion. Excess fibronectin was then removed by rinses with PBS.

**Viability assay**

To preliminarily evaluate the substrate cytotoxicity, alamarBlue proliferation assay, that allows a continuous monitoring of cells in culture, was performed on HEK cells. Briefly, the alamarBlue Reagent (Invitrogen DAL 1100) was diluted 11:1 with DMEM without phenol red. 500 µl of the obtained solution were added on each well. The solution was incubated for 3 hours at physiological condition. Once removed from the well, the emission at 590 nm of three aliquots per each well was measured. The emission value was acquired 5 times per each aliquot to obtain a reliable measure. Furthermore, the biocompatibility of the patterning process was accomplished on the more relevant biological model C2C12 via HOECHST 33342/NucGreen Dead 488 ReadyProbes assay. The substrates were incubated in extracellular containing the two dyes (HOECHST 33342 (10 µg ml$^{-1}$) and NucGreen Dead 488 ReadyProbes Reagent (2 drops ml$^{-1}$)) for 5 minutes protected from ambient light. The samples were then washed with extracellular solution and multiple images were acquired with a Nikon Eclipse Ti-S epifluorescence inverted microscope. Standard DAPI and FITC filter sets were employed for HOECHST and NucGreen respectively. The percentage of viable cells was estimated by counting the total number of cells nuclei (stained by HOECHST) and the total number of dead cells nuclei (stained by NucGreen).

**Scanning electron microscopy (SEM)**

C2C12 cells were plated on Patterned HDPE-P3HT (97%-3% in volume) substrates. These substrates were prepared for SEM following three steps. First, fixation in glutaraldehyde 4% in PBS for 20 minutes at room temperature; Second, immersion in increasing concentrations of ethanol (20%, 30%, 40%, 50%, 60%, 70%, 80%, 90% and 100%, 20 minutes for each concentration) followed by air-drying at room temperature; Three, evaporation of a thin gold layer on top of samples surface (thickness 5nm, 1.5 Cr adhesion layer).

All SEM images were acquired by using a TESCAN MIRA III scanning electron microscope (operating voltage 3 kV, working distance 7 mm, stage tilt angle 0°).

**Electrophysiology**

Standard patch clamp recordings were performed using an Axopatch 200B (Axon Instruments) coupled to an inverted microscope (Nikon Eclipse Ti). HEK and C2C12 cells seeded on bare glass/HDPE/HDPE:P3HT blend were measured in whole-cell configuration with freshly pulled glass pipettes (4-7 MΩ), filled with the following intracellular solution [mM]: 12 KCl, 125 K-Gluconate, 1MgCl2, 0.1 CaCl2, 10 EGTA, 10 HEPES, 10 ATP-Na2.

The extracellular solution contained [mM]: 135 NaCl, 5.4 KCl, 5 HEPES, 10 Glucose, 1.8 CaCl2, 1 MgCl2.

Only single cells were selected for recordings. Acquisition was performed with pClamp-10 software

(Axon Instruments). Membrane currents were low pass filtered at 2 kHz and digitized with a sampling rate of 10 kHz (Digidata 1440 A, Molecular Devices). Data were analyzed with Origin 9.0 (OriginLab Corporation) and with Matlab software.

The light source for excitation was provided by a green LED coupled to the fluorescence port of the microscope and characterized by maximum emission wavelength at 530 nm; the illuminated spot on the sample has an area of 0.23 mm$^2$ and a photoexcitation density of 40 mW/mm$^2$, as measured at the output of the microscope objective (Pobj).

**Cell Capacitance Measurement**

A double sinusoidal voltage-clamp signal is applied to the cell in whole-cell configuration. The response current signal is acquired and membrane capacitance, membrane resistance and access resistance are then extracted fitting the current with a custom Matlab program.

**Statistical analysis**

Data are represented as mean ± standard error of the mean (s.e.m.). Normal distribution was assessed using D'Agostino-Pearson's normality test. To compare two sample groups, either the Student's t-test or the Mann-Whitney U-test was used.. In all experimental settings, *P<0.05, **P<0.01, ***P<0.001, ***P<0.0001. Statistical analysis was performed using GraphPad Prism 7 software (GraphPad software, Inc, La Jolla, CA).

**Acknowledgements:** V. Vurro and A.D. Scaccabarozzi contributed equally to this work.

F. Lodola current affiliation is: Department of Biotechnology and Biosciences, Università di Milano-Bicocca, Piazza della Scienza, 2, 20126, Milano, Italy.

A.D. Scaccabarozzi current affiliation is: King Abdullah University of Science and Technology (KAUST) Solar Center (KSC), Division of Physical Sciences and Engineering, King Abdullah University of Science and Technology (KAUST), Thuwal, 23955-6900, Saudi Arabia.

The authors gratefully thank also Giulia Cavenago for her assistance with the viability measure.

**Funding:** A. S. and M. C. acknowledge the financial support of the European Research Council (ERC) under the European Union's Horizon 2020 research and innovation programme "HEROIC",

grant agreement 638059.

FM, AMR and FS acknowledge the financial support of the European Research Council (ERC) under the European Union's Horizon 2020 research and innovation programme "PAIDEIA", grant agreement 816313.

GMP thanks Fondazione Cariplo (grant n° 2018-0979) for the financial support.

**Authors contributions:** VV and GL planned the experiments. VV and ADS developed the fabrication process of the blend. VV and FL carried out photostimulation experiment. VV took care for cell cultures, and S.E.M sample preparation. AMR acquired the transient absorption data. FS design, developed and performed the laser ablation. FM acquired the S.E.M. images. VV, ADS, FL and GL wrote the main manuscript. All authors contributed to data interpretation and approved the final manuscript.

**Competing interests:** The authors declare that they have no competing interests.

## References

[1]  K. Deisseroth, *Nat Methods* **2011**, *8*, 26.

[2]  S. M. Ford, M. Watanabe, M. W. Jenkins, *J. Neural Eng.* **2018**, *15*, 011001.

[3]  F. Di Maria, F. Lodola, E. Zucchetti, F. Benfenati, G. Lanzani, *Chem. Soc. Rev.* **2018**, *47*, 4757.

[4]  A. Marino, S. Arai, Y. Hou, A. Degl'Innocenti, V. Cappello, B. Mazzolai, Y.-T. Chang, V. Mattoli, M. Suzuki, G. Ciofani, *ACS Nano* **2017**, *11*, 2494.

[5]  J. L. Carvalho-de-Souza, J. S. Treger, B. Dang, S. B. H. Kent, D. R. Pepperberg, F. Bezanilla, *Neuron* **2015**, *86*, 207.

[6]  J. Suzurikawa, H. Takahashi, R. Kanzaki, M. Nakao, Y. Takayama, Y. Jimbo, *Appl. Phys. Lett.* **2007**, *90*, 093901.

[7]  A. Starovoytov, J. Choi, H. S. Seung, *Journal of Neurophysiology* **2005**, *93*, 1090.

[8]  M. A. Colicos, B. E. Collins, M. J. Sailor, Y. Goda, *Cell* **2001**, *107*, 605.

[9]  R. Parameswaran, J. L. Carvalho-de-Souza, Y. Jiang, M. J. Burke, J. F. Zimmerman, K. Koehler, A. W. Phillips, J. Yi, E. J. Adams, F. Bezanilla, B. Tian, *Nature Nanotech* **2018**, *13*, 260.

[10]   R. Parameswaran, K. Koehler, M. Y. Rotenberg, M. J. Burke, J. Kim, K.-Y. Jeong, B. Hissa, M. D. Paul, K. Moreno, N. Sarma, T. Hayes, E. Sudzilovsky, H.-G. Park, B. Tian, *Proc Natl Acad Sci USA* **2019**, *116*, 413.

[11]   M. Zangoli, F. Di Maria, E. Zucchetti, C. Bossio, M. R. Antognazza, G. Lanzani, R. Mazzaro, F. Corticelli, M. Baroncini, G. Barbarella, *Nanoscale* **2017**, *9*, 9202.

[12]   E. Zucchetti, M. Zangoli, I. Bargigia, C. Bossio, F. Di Maria, G. Barbarella, C. D'Andrea, G. Lanzani, M. R. Antognazza, *J. Mater. Chem. B* **2017**, *5*, 565.

[13]   F. Lodola, V. Vurro, S. Crasto, E. Di Pasquale, G. Lanzani, *Adv. Healthcare Mater.* **2019**, *8*, 1900198.

[14]   D. Rand, M. Jakešová, G. Lubin, I. Vėbraitė, M. David-Pur, V. Đerek, T. Cramer, N. S. Sariciftci, Y. Hanein, E. D. Głowacki, *Adv. Mater.* **2018**, *30*, 1707292.

[15]   L. Bareket, N. Waiskopf, D. Rand, G. Lubin, M. David-Pur, J. Ben-Dov, S. Roy, C. Eleftheriou, E. Sernagor, O. Cheshnovsky, U. Banin, Y. Hanein, *Nano Lett.* **2014**, *14*, 6685.

[16]   J. Hopkins, L. Travaglini, A. Lauto, T. Cramer, B. Fraboni, J. Seidel, D. Mawad, *Adv. Mater. Technol.* **2019**, *4*, 1800744.

[17]   A. Savchenko, V. Cherkas, C. Liu, G. B. Braun, A. Kleschevnikov, Y. I. Miller, E. Molokanova, *Sci. Adv.* **2018**, *4*, eaat0351.

[18]   G. M. Paternò, E. Colombo, V. Vurro, F. Lodola, S. Cimò, V. Sesti, E. Molotokaite, M. Bramini, L. Ganzer, D. Fazzi, C. D'Andrea, F. Benfenati, C. Bertarelli, G. Lanzani, *Adv. Sci.* **2020**, *7*, 1903241.

[19]   M. L. DiFrancesco, F. Lodola, E. Colombo, L. Maragliano, M. Bramini, G. M. Paternò, P. Baldelli, M. D. Serra, L. Lunelli, M. Marchioretto, G. Grasselli, S. Cimò, L. Colella, D. Fazzi, F. Ortica, V. Vurro, C. G. Eleftheriou, D. Shmal, J. F. Maya-Vetencourt, C. Bertarelli, G. Lanzani, F. Benfenati, *Nat. Nanotechnol.* **2020**, *15*, 296.

[20]   J. Rivnay, R. M. Owens, G. G. Malliaras, *Chem. Mater.* **2014**, *26*, 679.

[21]   S. Zhang, E. Hubis, G. Tomasello, G. Soliveri, P. Kumar, F. Cicoira, *Chem. Mater.* **2017**, *29*, 3126.

[22]   M. Ermis, E. Antmen, V. Hasirci, *Bioactive Materials* **2018**, *3*, 355.

[23]   C. O. Chantre, S. P. Hoerstrup, K. K. Parker, *Current Opinion in Biomedical Engineering* **2019**, *10*, 97.

[24]   T. Fujie, A. Desii, L. Ventrelli, B. Mazzolai, V. Mattoli, *Biomed Microdevices* **2012**, *14*, 1069.

[25]   E. A. Roth, T. Xu, M. Das, C. Gregory, J. J. Hickman, T. Boland, *Biomaterials* **2004**, *25*, 3707.

[26]   A. W. Feinberg, A. Feigel, S. S. Shevkoplyas, S. Sheehy, G. M. Whitesides, K. K. Parker, *Science* **2007**, *317*, 1366.

[27]   A. W. Feinberg, P. W. Alford, H. Jin, C. M. Ripplinger, A. A. Werdich, S. P. Sheehy, A. Grosberg, K. K. Parker, *Biomaterials* **2012**, *33*, 5732.

[28]   D. Kai, M. P. Prabhakaran, G. Jin, S. Ramakrishna, *J. Biomed. Mater. Res.* **2011**, *98B*, 379.

[29]   Y. Orlova, N. Magome, L. Liu, Y. Chen, K. Agladze, *Biomaterials* **2011**, *32*, 5615.

[30]   L. F. Deravi, N. R. Sinatra, C. O. Chantre, A. P. Nesmith, H. Yuan, S. K. Deravi, J. A. Goss, L. A. MacQueen, M. R. Badrossamy, G. M. Gonzalez, M. D. Phillips, K. K. Parker, *Macromol. Mater. Eng.* **2017**, *302*, 1600404.

[31]   Z. Wang, D. Tonderys, S. E. Leggett, E. K. Williams, M. T. Kiani, R. Spitz Steinberg, Y. Qiu, I. Y. Wong, R. H. Hurt, *Carbon* **2016**, *97*, 14.

[32]   A. Curtis, C. Wilkinson, *Biomaterials* **1997**, *18*, 1573.

[33]   T. Guo, J. P. Ringel, C. G. Lim, L. G. Bracaglia, M. Noshin, H. B. Baker, D. A. Powell, J. P. Fisher, *J. Biomed. Mater. Res.* **2018**, *106*, 2190.

[34]   A. Chen, D. K. Lieu, L. Freschauf, V. Lew, H. Sharma, J. Wang, D. Nguyen, I. Karakikes, R. J. Hajjar, A. Gopinathan, E. Botvinick, C. C. Fowlkes, R. A. Li, M. Khine, *Adv. Mater.* **2011**, *23*, 5785.

[35]   F. Greco, T. Fujie, L. Ricotti, S. Taccola, B. Mazzolai, V. Mattoli, *ACS Appl. Mater. Interfaces* **2013**, *5*, 573.

[36]   J. U. Lind, T. A. Busbee, A. D. Valentine, F. S. Pasqualini, H. Yuan, M. Yadid, S.-J. Park, A. Kotikian, A. P. Nesmith, P. H. Campbell, J. J. Vlassak, J. A. Lewis, K. K. Parker, *Nature Mater* **2017**, *16*, 303.

[37]   L. Ricotti, B. Trimmer, A. W. Feinberg, R. Raman, K. K. Parker, R. Bashir, M. Sitti, S. Martel, P. Dario, A. Menciassi, *Sci. Robot.* **2017**, *2*, eaaq0495.

[38]   C. Müller, S. Goffri, D. W. Breiby, J. W. Andreasen, H. D. Chanzy, R. A. J. Janssen, M. M. Nielsen, C. P. Radano, H. Sirringhaus, P. Smith, N. Stingelin-Stutzmann, *Adv. Funct. Mater.* **2007**, *17*, 2674.

[39]   K. McMAHON, A. W. Anderson, J. Blair, E. Oakeley, N. Malouf, 8.

[40]   S. Goffri, C. Müller, N. Stingelin-Stutzmann, D. W. Breiby, C. P. Radano, J. W. Andreasen, R. Thompson, R. A. J. Janssen, M. M. Nielsen, P. Smith, H. Sirringhaus, *Nature Mater* **2006**, *5*, 950.

[41]   A. Kumar, M. A. Baklar, K. Scott, T. Kreouzis, N. Stingelin-Stutzmann, *Adv. Mater.* **2009**, *21*, 4447.

[42]   T. A. M. Ferenczi, C. Müller, D. D. C. Bradley, P. Smith, J. Nelson, N. Stingelin, *Adv. Mater.* **2011**, *23*, 4093.

[43]   A. D. Scaccabarozzi, N. Stingelin, *J. Mater. Chem. A* **2014**, *2*, 10818.

[44]   O. P. Dimitriev, *Nanoscale Res Lett* **2017**, *12*, 510.

[45]   S. Riera-Galindo, F. Leonardi, R. Pfattner, M. Mas-Torrent, *Adv. Mater. Technol.* **2019**, *4*, 1900104.

[46]   A. C. Arias, F. Endicott, R. A. Street, *Adv. Mater.* **2006**, *18*, 2900.

[47]   A. Campos, S. Riera-Galindo, J. Puigdollers, M. Mas-Torrent, *ACS Appl. Mater. Interfaces* **2018**, *10*, 15952.

[48]   C. Hellmann, F. Paquin, N. D. Treat, A. Bruno, L. X. Reynolds, S. A. Haque, P. N. Stavrinou, C. Silva, N. Stingelin, *Adv. Mater.* **2013**, *25*, 4906.

[49]   C. Hellmann, N. D. Treat, A. D. Scaccabarozzi, J. Razzell Hollis, F. D. Fleischli, J. H. Bannock, J. de Mello, J. J. Michels, J.-S. Kim, N. Stingelin, *J. Polym. Sci. Part B: Polym. Phys.* **2015**, *53*, 304.

[50]     D. Abbaszadeh, A. Kunz, G. A. H. Wetzelaer, J. J. Michels, N. I. Crăciun, K. Koynov, I. Lieberwirth, P. W. M. Blom, *Nature Mater* **2016**, *15*, 628.

[51]     C. Müller, T. A. M. Ferenczi, M. Campoy-Quiles, J. M. Frost, D. D. C. Bradley, P. Smith, N. Stingelin-Stutzmann, J. Nelson, *Adv. Mater.* **2008**, *20*, 3510.

[52]     G. Lanzani, G. Cerullo, S. Stagira, S. D. Silvestri, F. Garnier, . . . 5.

[53]     P. J. Brown, H. Sirringhaus, M. Harrison, M. Shkunov, R. H. Friend, *Phys. Rev. B* **2001**, *63*, 125204.

[54]     S. Cook, A. Furube, R. Katoh, *Energy Environ. Sci.* **2008**, *1*, 294.

[55]     E. W. Snedden, A. P. Monkman, F. B. Dias, *J. Phys. Chem. C* **2012**, *116*, 86.

[56]     N. Banerji, S. Cowan, E. Vauthey, A. J. Heeger, *J. Phys. Chem. C* **2011**, *115*, 9726.

[57]     F. C. Spano, *The Journal of Chemical Physics* **2005**, *122*, 234701.

[58]     J. Clark, C. Silva, R. H. Friend, F. C. Spano, *Phys. Rev. Lett.* **2007**, *98*, 206406.

[59]     I. Bargigia, E. Zucchetti, A. R. S. Kandada, M. Moreira, C. Bossio, W. P. D. Wong, P. B. Miranda, P. Decuzzi, C. Soci, C. D'Andrea, G. Lanzani, *ChemBioChem* **2019**, *20*, 532.

[60]     N. Martino, P. Feyen, M. Porro, C. Bossio, E. Zucchetti, D. Ghezzi, F. Benfenati, G. Lanzani, M. R. Antognazza, *Sci Rep* **2015**, *5*, 8911.

[61]     F. Lodola, N. Martino, G. Tullii, G. Lanzani, M. R. Antognazza, *Sci Rep* **2017**, *7*, 8477.

# Supporting Information

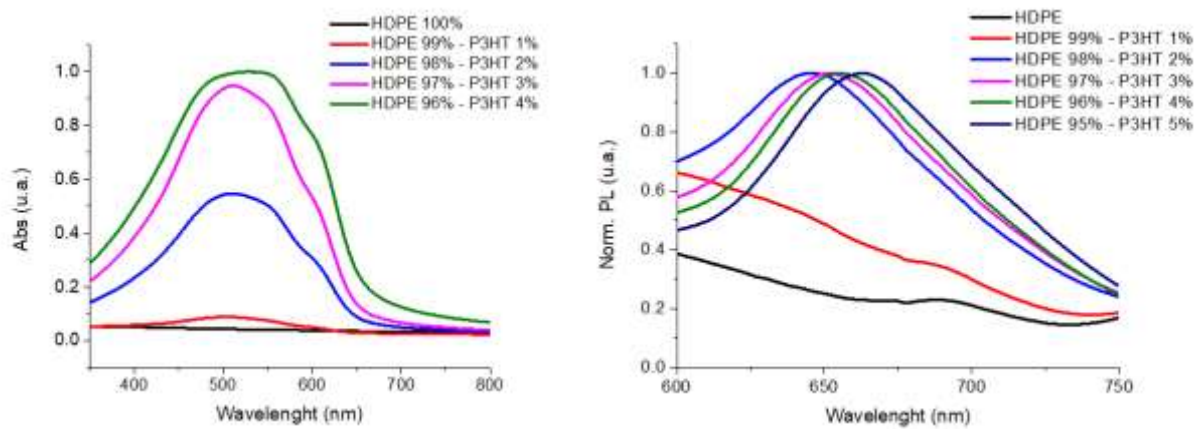

*Figure S1. Absorbance and Emission Comparison.* On the left the absorbance spectra of the P3HT-Tf and of Blend film obtained with different percentage of P3HT (Blend (1% P3HT), Blend (2% P3HT), Blend (3% P3HT) and Blend (4% P3HT)). The 5% P3HT Blend is not reported because of the saturation of the detector. These absorbance spectra are normalized by the maximum of the 4% Blend. On the right, the normalized emission spectra of the same films are reported.

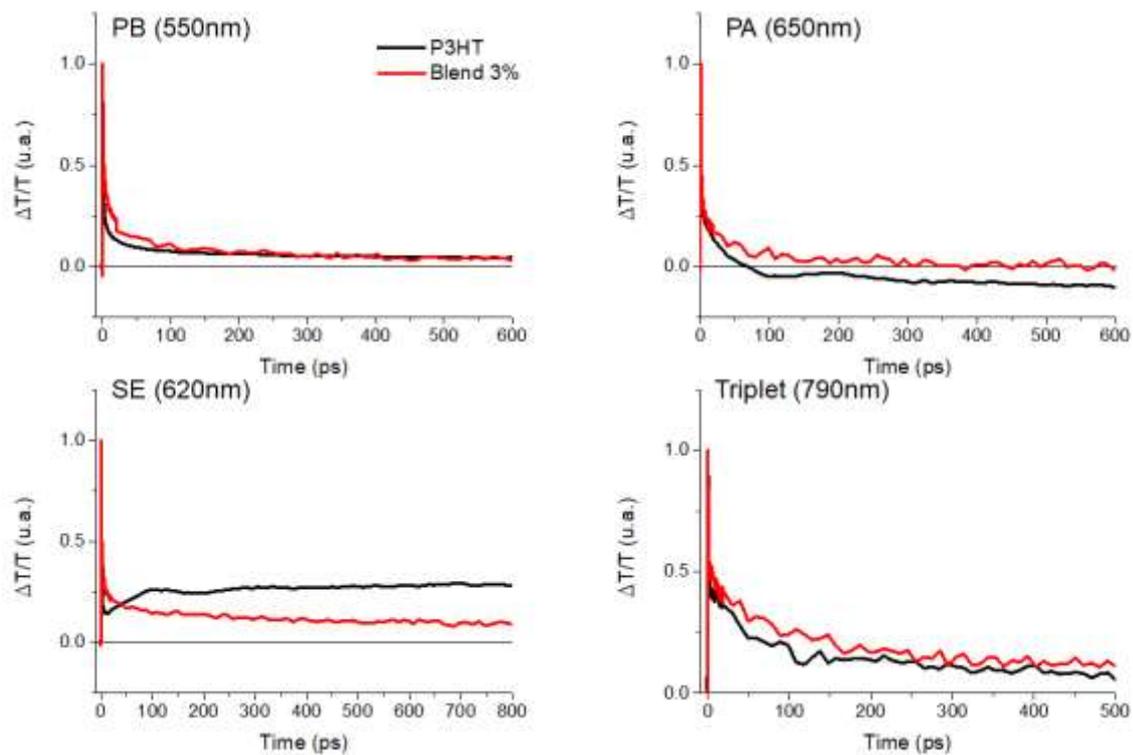

*Figure S2. P3HT-TF and Blend (3% P3HT) kinetic decay comparison.* The graphs show the decay of the Photo Bleaching (PB), Photoinduced Absorption (PA), Stimulated Emission (SE) and Triplet-Triplet Absorption (Triplet) for the P3HT-TF and for the Blended film (with 3% of P3HT). Data are obtained with excitation at 530 nm. The decay was fitted with a single exponential decay. The extracted characteristic times for P3HT-TF are $\tau_{PB} = 4.48\ ps$, $\tau_{PA} = 12.74\ ps$ and $\tau_{SE} = 153.69\ ps$ while for the Blend 3% $\tau_{PB} = 11.14\ ps$, $\tau_{PA} = 9.91\ ps$ and $\tau_{SE} = 8.17\ ps$.

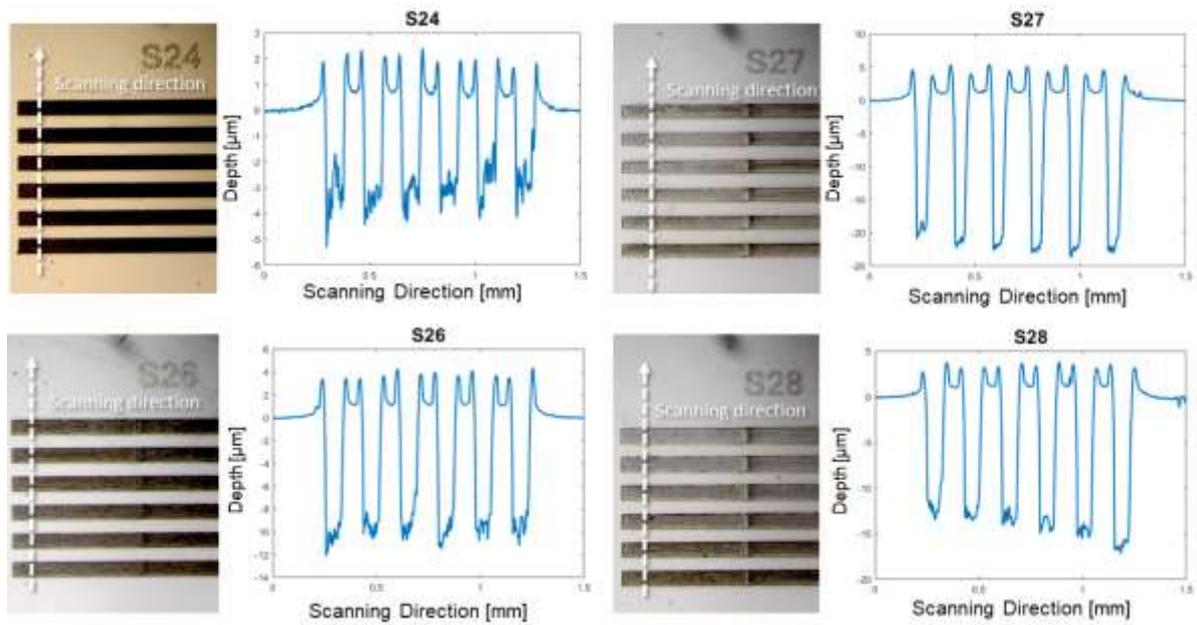

*Figure S3. Pattern Depth Tuning ability.* Patterns with different depths obtained upon changing the laser ablation parameters. In particular, the passage number of the laser on the substrate is changed. In the S28 pattern each line has a different number of ablations, this effect results in a different depth of each line.

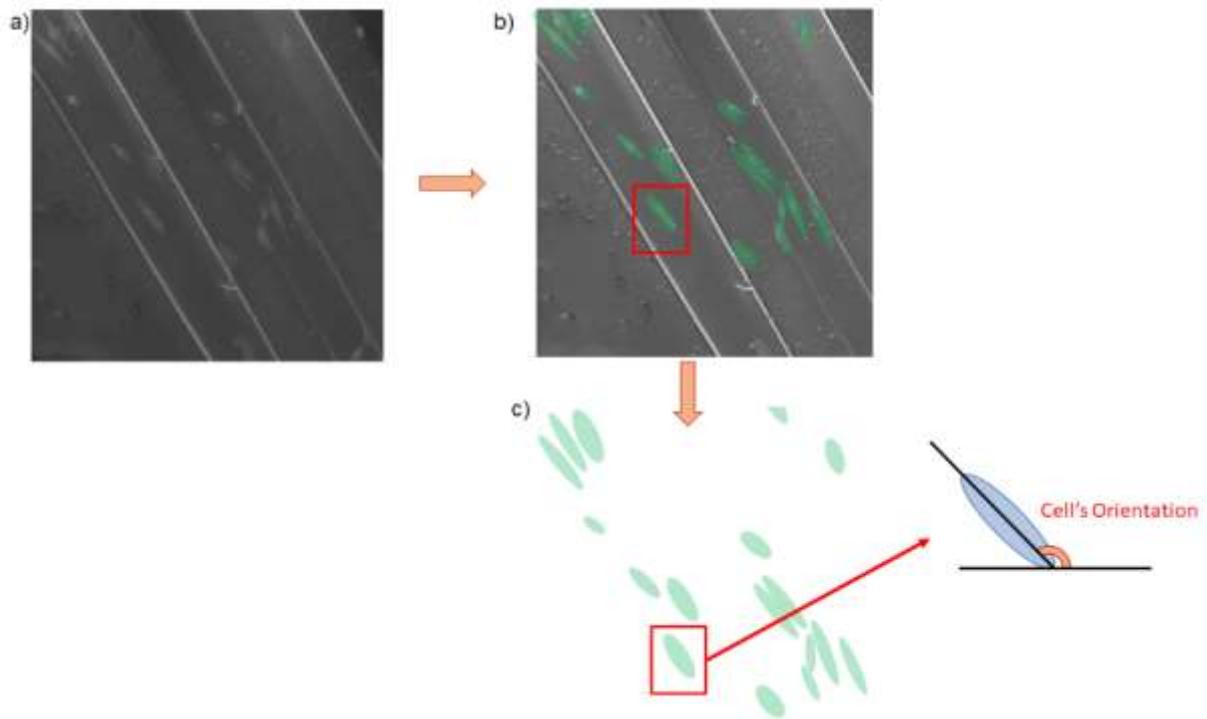

*Figure S4. Cell Orientation Extraction.* The cell orientation is measured by modelling each cells as an ellipse. The process starts from the SEM image (a). The cell is detected and associated to an ellipse (b). The angle formed by the x axis and the major axis of the ellipse is measured and associated to the direction of that specific cell (c). All measured directions were then normalized by subtracting the pattern direction (angle formed between the pattern and the x axis) in order to compare images acquired with different preferential orientations and associate the zero angle with those cells perfectly aligned with pattern direction.

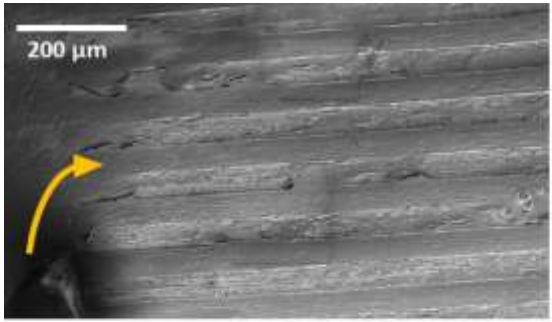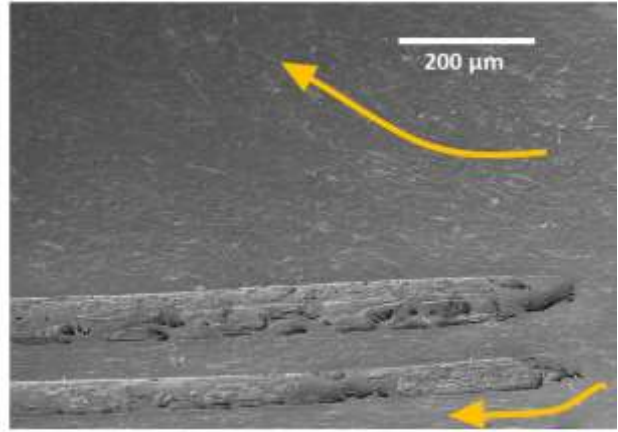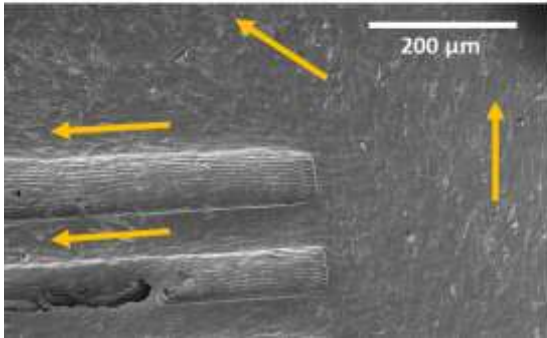

*Figure S5. High Cell Density Alignment Effect.* SEM images show the influence of the pattern on the cell directionality.